# Model error and its estimation, with particular application to loss reserving

By Greg Taylor

School of Risk and Actuarial Studies, University of New South Wales, Randwick, AUSTRALIA

and

Gráinne McGuire

Taylor Fry, Sydney, AUSTRALIA

**Abstract.** This paper is concerned with forecast error, particularly in relation to loss reserving. This is generally regarded as consisting of three components, namely parameter, process and model errors. The first two of these components, and their estimation, are well understood, but less so model error. Model error itself is considered in two parts: one part that is capable of estimation from past data (internal model error), and another part that is not (external model error). Attention is focused here on internal model error.

Estimation of this error component is approached by means of Bayesian model averaging, using the Bayesian interpretation of the LASSO. This is used to generate a set of admissible models, each with its prior probability and the likelihood of observed data. A posterior on the model set, conditional on the data, results, and an estimate of model error (contained in a loss reserve) is obtained as the variance of the loss reserve according to this posterior.

The population of models entering materially into the support of the posterior may turn out to be "thinner" than desired, and bootstrapping of the LASSO is used to gain bulk. This provides the bonus of an estimate of parameter error also. It turns out that the estimates of parameter and model errors are entangled, and dissociation of them is at least difficult, and possibly not even meaningful. These matters are discussed.

The majority of the discussion applies to forecasting generally, but numerical illustration of the concepts is given in relation to insurance data and the problem of insurance loss reserving.

# 1. Introduction

This paper is concerned with **forecast error**, particularly in relation to **loss reserving**. The majority of the concepts are fairly general, and might be applied in many forecasting environments, but the illustrations given here relate to loss reserving. To an extent, the paper is a sequel to McGuire, Taylor and Miller (2021), which dealt with estimation of a loss reserve by means of the LASSO. Both papers make applications of the LASSO, the first paper to point estimation, and the sequel to estimation of forecast error, particularly model error.

When a forecast is made on the basis of a model of observations, it will inevitably contain an error relative to the true value yet to be observed. Estimation of the properties of this error will provide some understanding of the reliability of the forecast. This has been an issue in the loss reserving literature since raised by Reid (1978), De Jong and Zehnwirth (1983) and Taylor and Ashe (1983).

In subsequent years, forecast error has been decomposed into a number of components, most notably parameter, process and model errors (see e.g. Taylor (1988), O'Dowd, Smith and Hardy (2005), Taylor and McGuire (2016), Taylor (2021)). Estimation of the first two of these three has become well understood, but there has been little development of the estimation of model error.

A notable exception was O'Dowd, Smith and Hardy (2005) and Risk Margins Task Force (2008), who provided a framework for the estimation of each component of forecast error using scorecards to score subjectively a range of factors identified as likely to influence the quantum of forecast error.

Where objective estimates are concerned, Taylor (2021) investigated model distribution error, a component of model error and Bignozzi and Tsanakas (2015) consider model risk in the context of VaR estimation, which is of course relevant to loss reserving, but loss reserving models as such are beyond their scope. Blanchet, Lam, Tang and Yuan (2019) estimate the effect of model error on a performance statistic in terms of the extrema of that measure over the set of admissible models.

However, to the authors' knowledge, there has been no other progress in the actuarial loss reserving literature.

In the meantime, the subject has been addressed elsewhere in the economics and finance literature. Useful general overviews are given by Glasserman and Xu (2014) and Schneider and Schweizer (2015) in the context of financial risk management. The approach of Blanchet, Lam, Tang and Yuan (2019) is similar to the latter.

Huang, Lam and Tang (2021) estimate the total of parameter error and model error (they call these data variability and procedural variability respectively) in relation to deep neural networks. Their approach equates more or less to bootstrapping, but where the replications are obtained by random variation of the network initialization rather than data re-sampling.

The literature gives certain prominence to Bayesian model averaging **("BMA")** (Raftery, 1995, 1996; Raftery, Madigan and Hoeting, 1997; Hoeting, Madigan, Raftery, and Volinsky,1999; Clyde and George, 2004; Clyde and Ivesen, 2013), and model confidence sets have also been considered (Hansen, Lunde and Nason, 2011, and others).

An example of this approach appears in the econometric literature in Loaiza-Maya, Martin and Frazier (2021), whose **focus Bayesian prediction** is similar to the Bayesian approach followed in the present paper, but with conditional likelihood replaced by a scoring rule. Martin, Loaiza-Maya, Maneesoonthorn, Frazier and Ramírez-Hassan (2021) give a non-Bayesian presentation of the same ideas.

This literature has been highly valuable at the fundamental conceptual level. However, the concepts are not easy to operationalize, and the much of the literature does not address loss reserving specifically.

The present paper endeavours to fill some of the literature gaps identified above. It uses the LASSO (Hastie, Tibshirani and Friedman, 2009) to populate abstract concepts, such as model set and its prior distribution, that occur in the more theoretical literature. Since the LASSO may also be used as the source of a loss reserving model, this creates a direct nexus between that model and the estimation of its model error.

The result is that, using the procedures described here, one may perform the following entire sequence of operations:

- model a data set of claim observations and extract a point estimate of loss reserve;
- move on to estimating the distributions of several components of forecast error that constitute a major part of the total;
- supplement these with the distributions of the missing components, derived from other sources;
- apply these distributions to the calculation of loss reserve risk margins, or any other quantities of interest that depend on the distribution of forecast error.

The paper is arranged as follows. After the establishment of the necessary mathematical framework and notation in Section 2, the structure of the problem to be considered is established with a review of the components of forecast error, and thereafter the paper focuses on one particular component, **internal model structure error ("IMSE")** (Section 3). Section 4 discusses the estimation of IMSE in the abstract, and the ingredients required for it. Then Section 5 puts these concepts to work in the specific context of the LASSO. This derives an estimate of the distribution of IMSE, but this estimate is then strengthened by bootstrapping the LASSO in Section 6. The whole procedure is then applied to several synthetic data sets, of varying complexity, in the derivation of numerical results in Section 7. Finally, Section 8 summarizes and considers the successes and limitations of the paper in the attainment of its objectives.

## 2. Reserving framework and notation

As far as possible, the notation here will follow that of McGuire, Taylor and Miller (2021). Accordingly, the analysis below will be concerned with the conventional **claim triangle**. Some random variable of interest $Y$ is labelled by **accident period** $i = 1,2,...,I$ and **development period** $j = 1,2,...,I - i + 1$. In this setup, a **cell** of the triangle refers to the combination $(i,j)$, and the observation $Y$ in this cell denoted $Y_{ij}$. The **payment period** to which cell $(i,j)$ relates will be denoted by $t = i + j - 1$.

Let $\Delta$ denote the collection of cells (i.e. ordered pairs $(i, j)$) of which the triangle consists, and let $\mathcal{Y}$ denote the observations on these cells, i.e. $\mathcal{Y} = \{Y_{ij} : (i, j) \in \Delta\}$. Accident and development periods will be assumed of equal duration, but not necessarily years. As further notation, $E[Y_{ij}] = \mu_{ij}, Var[Y_{ij}] = \sigma_{ij}^2$. A realization of $Y_{ij}$ will be denoted $y_{ij}$.

Let $Z_{ij}$ be any random vector defined on $\Delta$ and $z_{ij}$ its realization. It will sometimes be useful to vectorize these quantities, and so $Z$ will denote the column vector of all $Z_{ij}$ listed in some defined order, and $z$ the corresponding vector of all $z_{ij}$.

This paper will be concerned with **forecasts** produced by models fitted to the data $Y$. Forecasts are made in respect of the $Y_{ij}$ for $(i, j) \in \Delta^*$, some set of cells, disjoint from $\Delta$, and relating to the future, i.e. $i + j > I + 1$. These $Y_{ij}$ will now be denoted $Y_{ij}^*$, and the vector of these will be $Y^*$.

The forecast of $Y_{ij}^*$ by model $M$ will be denoted $\hat{Y}_{ij}^*(M)$ and is equal to $f_M(Y; i, j)$ for some real-valued function $f_M$. As the notation indicates the forecast is $M$-dependent but, as this notation is cumbersome, the $M$ will be suppressed and the forecast written as simply $\hat{Y}_{ij}^*$ when this does not create any ambiguity. Other quantities derived from $\hat{Y}_{ij}^*$ (e.g. immediately below) will also be notated without explicit mention of $M$.

A model $M$ will include a likelihood function $L(.|M)$ for the observations so that the likelihood of the data vector $Y$ is $L(Y|M)$. Let $\ell(Y|M) = \ln L(Y|M)$, the log-likelihood function.

It will be convenient to denote a value fitted by the model to a past observation $y_{ij}$ by $\hat{\mu}_{ij} = f_M(Y; i, j)$. According to the vector notation given above, $\hat{\mu}$ denotes the vector of $\hat{\mu}_{ij}$ for $(i, j) \in \Delta$. Similarly, let $\hat{Y}^*$ denote the vector of $\hat{Y}_{ij}$ for $(i, j) \in \Delta^*$.

The modeller may select the model $M$ but, in practice, will rarely know whether it is a correct representation of the data. Therefore, assume that $M$ is selected from some collection $\mathcal{M}$ of candidate models, hereafter called the **model set**. Assume further that the model set is equipped with a measure $\pi$.

Suppose that $R = g(Y^*)$, for some real-valued function $g$, and define its forecast as $\hat{R} = g(\hat{Y}^*)$. An example is $R = 1^T Y^*$, where 1 is a vector of the same dimension as $Y^*$ and with all components equal to unity. If the $Y_{ij}^*$ denote claim payments, then $R$ is the amount of outstanding claim liability.

The **forecast error** associated with forecast $\hat{R}$ will be defined as

$$e = R - \hat{R}. \tag{2.1}$$

Later sections will make use of **open-ended ramp functions**. These are single-knot linear splines with zero gradient in the left-hand segment and unit gradient in the right-hand segment. In a machine learning context, one of these would be referred to as a **rectified linear unit ("ReLu")**. Let $R_K(x)$ denote the open-ended ramp function with knot at $K$. Then

$$R_K(x) = max(0, x - K). \tag{2.2}$$

For a given condition $c$, define the **indicator function** $I(c) = 1$ when $c$ is true, and $I(c) = 0$ when $c$ is false. Further, define the **Heaviside function** $H_k(x) = I(x \geq k)$.

# 3. Components of forecast error

Some regulatory regimes require that a **capital margin** be associated with a technical reserve such that the total of reserve plus margin equal at least the **Value at Risk ("VaR")** of the liability at some high percentile, such as 99.5%. These regimes may also require the evaluation of a **risk margin** within the capital margin. For example, the Australian prudential standards require a loss reserve to be at least equal to the 75% VaR of the associated liability (Australian Prudential Regulatory Authority, 2018).

IFRS17 requires a risk adjustment for non-financial risk. While there is no prescribed method for calculation, margins based on the **Value at Risk ("VaR")** are likely to be widely used.

For the sake of definiteness, this paper proceeds on the basis that the specific liability forecast under consideration is the loss reserve. For some regulatory regimes (e.g. IFRS17) it might be the one-year-ahead forecast of claim payments, with an associated VaR. In such cases, the model error methodology set out in subsequent sections translates readily to this alternative situation.

The requirement of a VaR necessitates the estimation of the **distribution** of a forecast liability, as opposed to a simple point estimate. Often, especially in the case of low- or medium-percentile VaRs, it is reasonable to assume that the required distribution is characterized by its mean (the point estimate) and variance. Hence the need for examination of the variance of forecast error.

It consists of a number of identifiable components. Decomposition of forecast error is discussed by Taylor (2000), Taylor and McGuire (2016), McGuire, Taylor and Miller (2021), Taylor (2021), O'Dowd, Smith and Hardy (2005), Risk Margins Task Force (2008) and Hastie, Tibshirani and Friedman (2009), Huang, Lam and Zhang (2021), among others.

The different authors use slightly different terminology, and so Table 3-1 displays the correspondences between the different terminologies. A blank entry indicates that the component concerned is not explicitly considered by the authors in question. Correspondences are sometimes exact, but at other times are a little rough because of different approaches taken by different authors.

**Table 3-1 Terminologies for components of forecast error**

| Taylor and co-authors | O'Dowd et al. | Risk Margins Task Force | Hastie et al. | Machine learning[a] |
|---|---|---|---|---|
| Model bias | | | Model bias (partial) | |
| Parameter error | Independent parameter risk | Independent parameter risk | Variance | Data variability |
| Process error | Independent process risk | Independent process risk | Irreducible error | Aleatoric uncertainty |
| Internal model structure error | Model specification risk; Systemic parameter risk | Internal systemic risk | Model bias (partial) | Procedural variability |
| External model structure error | Future systemic risk | External systemic risk | Model bias (partial) | |
| Model distribution error | | | | |

**Note:** [a] Huang, Lam and Zhang (2021) bundle all model error other than aleatoric under the term "epistemic uncertainty".

The most comprehensive decomposition is given by Taylor (2021), on which the following is largely based. Henceforth, let the model $M$ introduced in Section 2 consist of just its distribution-free part, i.e. the algebraic form of its non-stochastic part (e.g. the linear predictor in the case of a GLM), and let $F$ denote the model distribution of the stochastic error.

Let $\mu, \hat{\mu}$ denote $E[R], E_{M,F}[\hat{R}|M,F]$ respectively. Then the forecast error (2.1) may be expressed as

$$e = (\mu - \hat{\mu}) + (R - \mu) - (\hat{R} - E[\hat{R}|M,F]) - (E[\hat{R}|M,F] - \hat{\mu}). \tag{3.1}$$

and it follows (Taylor, 2021) that the **mean square error of prediction (MSEP)** of $\hat{R}$ is

$$\begin{aligned} MSEP[e] &= E[e^2|M,F] \\ &= (\mu - \hat{\mu})^2 + E_F Var[R|F] + E_{M,F} Var[\hat{R}|M,F] \\ &\quad + Var_{M,F} E[\hat{R}|M,F] \end{aligned} \tag{3.2}$$

With a slight abuse of notation, the quantity $MSEP[e]$ will also be referred to as forecast error when this usage is unambiguous. The four terms on the right may be recognised as respectively model bias, process error, parameter error **("PaE")** and model error.

The model error may be further decomposed with a small amount of manipulation:

$$Var_{M,F} E[\hat{R}|M,F] = E_F Var_M E[\hat{R}|M,F] + Var_F E_M E[\hat{R}|M,F]. \tag{3.3}$$

The two members on the right are labelled **model structure error** and **model distribution error** in Taylor (2021). If the latter is assumed away by assuming the distribution of $F$ to be concentrated in a single point, then (3.3) collapses to just

$$Var_{M,F} E[\hat{R}|M,F] = Var_M E[\hat{R}|M]. \tag{3.4}$$

O'Dowd, Smith and Hardy (2005) recognize that some of the variation over models $M \in \mathcal{M}$ relates to mis-specification of the model fitted to past data, and part to the extrapolation of that model in its forecasts.

In the case of a GLM, for example, the data will have been modelled using some linear response $X\beta$, and the data set augmented by the forecast experience will use an augmented linear response $X^+\beta^+$, where

$$X^+ = \begin{bmatrix} X & 0 \\ X^* & X^{**} \end{bmatrix}, \beta^+ = \begin{bmatrix} \beta \\ \beta^* \end{bmatrix}, \tag{3.5}$$

and the parameter sub-vector $\beta^*$ is not estimated from data, but consists of assumptions about the future, which are combined with the design matrix subcomponent $X^{**}$ to form the augmented part of the linear predictor. For example, superimposed inflation (**"SI"**), a diagonal effect, might be estimated at one rate in past data, but a different rate assumed for the future. The past rate would be included in $\beta$, and the future in $\beta^*$.

The forecast of future claims experience depends on the linear response $X^*\beta + X^{**}\beta^*$, in which the first of the two members relates to past observations, and the second to assumptions about the future. Forecast errors in relation to these two are referred to as **internal** and **external model errors** respectively.

Hence, the first of the two members on the right side of (3.3) can be decomposed further:

$$E_F Var_M E[\hat{R}|M,F] = E_F Var_{M^{int}} E[\hat{R}|M,F] + E_F Var_{M^{ext}} E[\hat{R}|M,F], \tag{3.6}$$

where $M^{int}, M^{ext}$ are those parts of model $M$ generating internal and external model errors.

Substitution of (3.3) and (3.6) into (3.2) yields the full decomposition depicted in Figure 3-1.

**Figure 3-1  Full decomposition of model error**

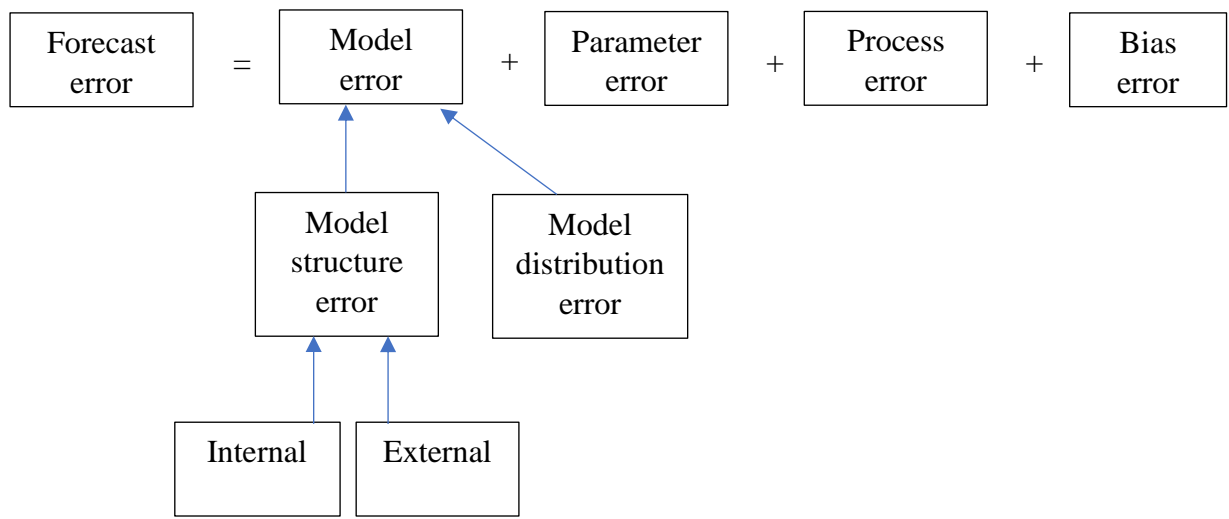

The estimation of parameter and process error is covered by England and Verrall (1999, 2001, 2002), Taylor (2000) and Taylor and McGuire (2016). Model distribution error is discussed by Taylor (2021).

The present paper will be concerned with model structure error, and specifically IMSE. As explained earlier in this section, this relates to any error in the selection of the algebraic form fitted to the data, and the objective here is to estimate this error on the basis of the data.

External model error, on the other hand, relates to future events and influences on future claim data and, by definition, there are no available data relevant to these. The estimation of this component of error must be performed by some means other than reference to past claim data (e,g, O'Dowd, Smith and Hardy, 2005), and is not addressed here.

## 4. Estimation of internal model structure error

### 4.1. Fundamental ingredients

The fundamentals of IMSE are discussed in Clyde and George (2004), Clyde and Iversen (2013), Hansen, Lunde and Nason (2011), Schneider and Schweizer (2015).

As mentioned in Section 2, one contemplates models $M \in \mathcal{M}$, the model set, with probability measure $\pi$ defined on $\mathcal{M}$. Each model $M$ defines a loss reserve estimate $\hat{R}(M) = 1^T \hat{Y}^*(M)$. With this machinery, one is in a position to calculate summary statistics of the estimated loss reserve over candidate models. Of particular interest are

$$E_{\mathcal{M}}[\hat{R}] = \int_{\mathcal{M}} \hat{R}(M) \, d\pi(M), \tag{4.1}$$

$$Var_{\mathcal{M}}[\hat{R}] = \int_{\mathcal{M}} \left[\hat{R}(M) - E_{\mathcal{M}}[\hat{R}]\right]^2 d\pi(M). \tag{4.2}$$

Quantity (4.2) measures the dispersion of estimated loss reserve over the entire model set, and will be of interest in the formulation of model error in Section 5.2.

### 4.2. Model confidence set

This is discussed by Hansen, Lunde and Nason (2011), Glasserman and Xu (2014) and Schneider and Schweizer (2015). In its most basic definition, a subset $\mathcal{S} \subset \mathcal{M}$ is a $100\alpha\%$ **model confidence set** relative to some reference model $M_0$ if $M_0 \in \mathcal{S}$ and $\pi(\mathcal{S}) = \alpha$. This is the approach of Hansen, Lunde and Nason (2011), but there are other possibilities. For example, Schneider and Schweizer (2015) are concerned with identification of the worst-case risk within a prescribed divergence radius of the reference model.

### 4.3. Bayesian model averaging

The concepts of BMA are discussed in Raftery (1995, 1996), Raftery, Madigan and Hoeting (1997), Hoeting, Madigan, Raftery, and Volinsky (1999), Clyde and George (2004), Clyde and Iversen (2013).

This approach uses the structure of Section 4.1 but with $\pi$ representing a Bayesian posterior distribution on $\mathcal{M}$. It is supposed that a prior distribution function $\pi_{pr}$ is specified on $\mathcal{M}$. Then application of Bayes theorem yields the posterior distribution

$$d\pi_{po}(M) = d\pi_{po}(M|Y = y) = \frac{L(y|M)d\pi_{pr}(M)}{\int_{M\in\mathcal{M}} L(y|M)d\pi_{pr}(M)}. \qquad (4.3)$$

Consider the case in which $\hat{R}(M)$ is any forecast of model $M$, not necessarily a loss reserve. Its posterior mean is

$$E[\hat{R}(M)|Y] = \int_{M\in\mathcal{M}} \hat{R}(M)d\pi_{po}(M), \qquad (4.4)$$

which is a weighted average of the forecasts across all models, with weights $d\pi_{po}(M)$. The concept of averaging across models was introduced to the loss reserving context by Taylor (1985).

# 5. LASSO estimation of internal model structure error and other forecast errors

## 5.1. The LASSO

### 5.1.1. Theoretical background

Consider a model $M$ of the data set $Y$ that takes the form

$$Y_{ij} \sim F(i, j; \beta, \theta), \qquad (5.1)$$

where $F$ is a distribution function (d.f.) defined within some family by parameter vectors $\beta, \theta$, and subject to

$$E[Y_{ij}] = f_M(Y; i, j, \beta). \qquad (5.2)$$

Thus, it is supposed that the mean $E[Y_{ij}]$ is defined by the parameters $\beta$, and the distribution around this mean defined by the parameters $\theta$.

A **LASSO regression** estimates $\beta$, of dimension $p$, as

$$\hat{\beta} = \underset{\beta}{\mathrm{argmin}} \left( \delta(Y, \hat{\mu}(\beta)) + \sum_{r=1}^{p} \lambda_r |\beta_r| \right), \qquad (5.3)$$

where the dependence of $\hat{\mu}$ on $\beta$ has been made explicit, $\delta$ is some measure of the separation between $Y$ and $\hat{\mu}(\beta)$, the $\beta_r$ are the components of $\beta$, and the $\lambda_r \geq 0, r = 1, \ldots, p$ are fixed values selected by the modeller.

A choice of $\delta$ such that the regression reduces to a GLM in the event that all $\lambda_r = 0$, is the negative log-likelihood function $\delta(Y, \hat{\mu}(\beta)) = -\ell(Y|M)$, in which case (5.3) becomes

$$\hat{\beta} = \underset{\beta}{\mathrm{argmin}} \left( -\ell(Y|\beta) + \sum_{r=1}^{p} \lambda_r |\beta_r| \right), \qquad (5.4)$$

where the model $M$ has been represented by its parameter vector $\beta$.

There are a couple of fundamental derivations of the estimator (5.3), specifically as (Hastie, Tibshirani and Friedman, 2009):

(a) the minimizer of the $\delta$ quantity subject to an upper bound on each $|\beta_r|$; and
(b) the maximizer of the *a posteriori* estimate of the $\delta$ quantity when each coefficient $\beta_r$ is random subject to a Laplace density.

Case (b) will be of interest here, and so is explained in a little more detail. It is assumed that $\beta_r$ is distributed with Laplace probability density

$$p(\beta_r) = \tfrac{1}{2}\lambda_r \exp(-\lambda_r|\beta_r|), \tag{5.5}$$

where $\lambda_r$ is a scale parameter. This is a symmetric density about a zero mean, and

$$Var(\beta_r) = 2/\lambda_r^2. \tag{5.6}$$

Then the posterior likelihood of $\beta$ is proportional to $exp\{-(-\ell(Y|\beta) + \lambda^T|\beta|)\}$, where the modulus operates on a vector element-wise and $\lambda$ denotes the column $p$-vector with the $\lambda_r$ as components. This posterior likelihood may be compared with (5.4).

### 5.1.2. Implementation

The application of the LASSO to data follows the procedure set out in McGuire, Taylor and Miller (2021). The target variable $Y$ denotes the observed claim payments over a $40 \times 40$ triangle. The model takes the same form as a GLM with log link, specifically supplementing (5.1) and (5.2) with

$$F(i,j;\beta,\theta) = Gamma(\mu_{ij}, \phi), \tag{5.7}$$

a Gamma d.f. with dispersion parameter $\phi$, and

$$\mu_{ij} = f_M(Y; i, j, \beta) = exp(X\beta). \tag{5.8}$$

In this formulation, the parameter $\theta$ from (5.1) is set equal to the $\phi$, the over-dispersion parameter. This parameter requires estimation, which is discussed later in this sub-section.

A LASSO can be implemented by means of the *R* package *glmnet* but only with Poisson ($\phi = 1$) rather than Gamma distribution. The solution is to do just this but, later, in the BMA of Section 5.2, adjust all Poisson distributions to Gamma with suitable $\phi$. A Gamma error term is preferred as more realistic than Poisson but, although the package *glmnet v4.0* purports to provide the means for this, numerical difficulties precluded its application here.

McGuire, Taylor and Miller (2021) propose that the covariates represented by design matrix $X$ be those of from a basis set that linearly spans a sufficiently extensive function space. The choice there, retained here, was

- ramp functions (2.2) $R_K(i), R_K(j), R_K(t), K = 0,1,\ldots,39$; and
- interactions of Heaviside functions $H_k(i)H_\ell(j), H_k(i)H_g(t), H_g(t)H_\ell(j), k, g, \ell = 2,3,\ldots,40$ for interactions.

All covariates were standardized, as described by McGuire, Taylor and Miller (2021). The orders of such basis sets are typically in the thousands or tens of thousands, depending on the dimensions of the data being used, but the LASSO will eliminate most of these from the model.

The vector $\lambda$ is restricted to the form $(0, \ldots, 0, \lambda', \lambda', \ldots, \lambda')^T$, where $\lambda' > 0$ and the leading zeros relate to specific data features that the modeller regards as of certain existence and wishes to force into the model – typically an intercept term is fitted without penalty. Thus, the terms of the linear response corresponding to these covariates are always included without penalty, and all other members of the linear response carry penalty parameters that are positive and equal. Details appear in McGuire, Taylor and Miller (2021).

A sequence of values of $\lambda'$ is chosen, covering a range from large to small. Let these be denoted by $\lambda'^{(q)}, q = 1, \ldots, Q$, and let $\lambda^{(q)}$ denote the vector $\left(0, \ldots, 0, \lambda'^{(q)}, \lambda'^{(q)}, \ldots, \lambda'^{(q)}\right)^T$, where $\lambda'^{(q)} \downarrow$ as $q \uparrow$. For each $q$, the LASSO is fitted to the data $Y$, i.e. $\beta$ is estimated according to (5.3), and the estimate denoted by $\hat{\beta}_{(q)}$. This induces a model $M_q$.

The fit of each $M_q$ to data is assessed by 8-fold cross validation using the Poisson deviance as the loss function. The cross-validation loss in the $s$-th fold is denoted by $CV_{qs}$. The average of the $CV_{qs}$ values across $s$ is denoted by $\overline{CV}_q$ and the standard error of the same $CV_{qs}$ values by $SE_q$.

Two models that are conventionally regarded as "optimal" in the literature are selected from the collection $\{M_q, q = 1, \ldots, Q\}$. These are:

- A "minCV" model $M_{q_{min}}$ such that $\overline{CV}_{q_{min}} = \min_{1 \leq q \leq Q} \overline{CV}_q$; and
- A "1se" model $M_{1se}$, where $M_{1se} = M_{q_{1se}}$ such that $q_{1se} = min\{q: \overline{CV}_q \leq \overline{CV}_{q_{min}} + SE_{q_{min}}\}$.

At this point, the parameter $\phi$ is estimated. The model structure (i.e. the covariates with non-zero coefficients) of the 1se model is extracted, and a GLM fitted to this structure. Henceforth, the preferred Gamma error is assumed, so the GLM is fitted using the *R* function *glm()* from the *stats* package with the same weights as in the LASSO fit. Since only a crude estimate of $\phi$ is yielded by *glm()*, the maximum likelihood estimate of $\phi$ is obtained from the *gamma.dispersion()* function from the *MASS* package.

This last value of $\phi$ is used in conjunction with an assumed Gamma error for all calculations henceforth. It is not re-estimated in the bootstrap replications of Section 6. The GLM itself is of no particular interest other than the estimation of $\phi$.

The 1se and minCV models generate the LASSO estimates of outstanding claim liability chosen by McGuire, Taylor and Miller (2021). The extrapolation of a model of past data to the future require decisions as to the extent to which any trends identified in the past are also extrapolated, and to what extent.

For this purpose, the decisions of McGuire, Taylor and Miller (2021) have been followed, in that all past trends have been extended into the future without modification. In more precise terms, if the linear response included in the model of past data is $f(i, j, t)$ (consistent of a linear combination of ramp functions and products of Heaviside functions), then the linear response for the future will also be taken as simply $f(i,j,t)$.

This has the benefit of even-handedness in the generation of a model set. There is no user intervention in the exclusion of models that fail to conform with preconceptions. On the other hand, however, it does admit some models that might be viewed as dubious. For example, any payment quarter effect that is modelled to take effect in the last few payment quarters of experience will be extended indefinitely into the future. Ultimately, such exotic models may need to be pruned away (Section 5.2.5).

### 5.1.3. Model bias

The LASSO's shrinkage of parameter estimates induces bias in model estimates. For this reason, the LASSO is sometimes used just for selection of the model form, and then the selected model refitted to the data without penalty, i.e. as a GLM. The result should be approximately unbiased.

McGuire, Taylor and Miller (2021) carried out some experimentation with this procedure in relation to the example data sets of Section 7.1, but commented that the results were not encouraging in that the refit caused a substantial deterioration in model performance in some cases.

See the further comment in Section 5.2.1.

### 5.2. Bayesian model averaging of LASSO

### 5.2.1. Generation of model set

The cell mean structure within the LASSO model is given by (5.2). Variation of $\beta$ there will create a multiplicity of models $\mathcal{M} = \{f_M(.;.,.,\beta), \beta \in \mathbb{R}^p\}$. It is of particular interest that, as $\lambda_r$ increases to a certain threshold, the LASSO forces $\beta_r$ to zero for that $\lambda_r$ and all greater values. This amounts to dropping a covariate from the model, equivalent to changing the cell mean's algebraic structure.

This is exactly one of the requirements of Section 4.1. In fact, the LASSO provides a rich model set that includes both variation in model algebraic structure and parametric variation within each distinct structure. The latter form of variation arises from variation in the penalty parameter, and is lost if a GLM is fitted to a LASSO model structure.

### 5.2.2. Model likelihood

As explained in Section 5.1.2, the process of BMA about to be described assumes a Gamma distribution for each observation $Y_{ij}$. The log-likelihood is

$$\ell(Y|\beta) = \sum_{(i,j)\in\Delta} \left[\gamma \ln c_{ij} - \ln \Gamma(\gamma) + (\gamma - 1)\ln y_{ij} - c_{ij}y_{ij}\right],$$

(5.9)

where $\gamma = 1/\phi$, $c_{ij} = 1/\phi\mu_{ij}$, with the $\mu_{ij}$ given by (5.8) and the estimation of $\phi$ is described in Section 5.1.2.

In the case of (5.8), $X$ consists of columns representing all functions in the basis set defined in Section 5.1.2. In the row corresponding to cell $(i,j)$, each of the functions is evaluated for those values of $i,j$.

### 5.2.3. Prior distributions of LASSO models
By (5.5),

$$p(\beta) = 2^{-p}\left(\prod_{r=1}^{p}\lambda_r\right)\exp(-\lambda^T|\beta|), \tag{5.10}$$

assuming the random parameters $\beta_r$ are independent, and this may be viewed as a prior distribution on the models $M \in \mathcal{M}$.

As appears in Section 5.1.2, the vector $\lambda$ is defined by the scalar $\lambda'$. It is of especial relevance that, in (5.10), $\lambda'$ functions as a dispersion parameter for the prior. There is a correspondence $q \Leftrightarrow \lambda'^{(q)}$, so that models may be labelled equivalently by $q$ or $\lambda'$. As noted in Section 5.1.2, models are numbered in such a way that $\lambda'^{(q)} \downarrow$ as $q \uparrow$. Decreasing $\lambda'^{(q)}$ implies lower penalty on parameters, and so model complexity increases steadily with increasing $q$. In the limit $\lambda'^{(q)} = 0$, the LASSO degenerates to a GLM with the set of covariates equal to the entire basis set, i.e. an extremely complex model.

Similarly, high values of $\lambda'^{(q)}$ force the model toward simplicity. As $\lambda'^{(q)} \to \infty$, the model tends to one containing only the unpenalized covariates.

Each value of $\lambda'^{(q)}$ also induces a posterior distribution $\pi_{po}\left(M_q|y;\lambda'^{(q)}\right)$, where the dependence on $q$ appears twice; first, the algebraic structure of the model $M_q$ depends on $q$ and, second, the posterior is influenced by the prior dispersion parameter. This sets up a sequence of relations $q \Leftrightarrow \lambda'^{(q)} \Rightarrow \pi_{po}^{(q)}$, where the last term is written as an abbreviation for $\pi_{po}\left(M_q|y;\lambda'^{(q)}\right)$.

Recall, however, that all of this applied in the context of Poisson likelihood (Section 5.1.2). Subsequently, this was replaced by a Gamma likelihood, with consequent change to the posterior. If $\pi_{po[G]}^{(q)}$ denotes the posterior that replaces $\pi_{po}^{(q)}$ when the Poisson likelihood is replaced by Gamma, then the last relation becomes $q \Leftrightarrow \lambda'^{(q)}_{[G]} \Rightarrow \pi_{po[G]}^{(q)}$, where $\lambda'^{(q)}_{[G]}$ is now the Laplace prior dispersion parameter to "equate" $\pi_{po[G]}^{(q)}$ to $\pi_{po}^{(q)}$, specifically to "equate" their modes. The quotes appear here to indicate that equality will usually be only approximate, due to discreteness of $q$.

The relation between $\lambda'^{(q)}$ and $\lambda'^{(q)}_{[G]}$ is unknown. In principle, it could be derived by analytic comparison of the posteriors $\pi_{po}^{(q)}$ and $\pi_{po[G]}^{(q)}$. Alternatively, and more simply, any required $\lambda'^{(q)}_{[G]}$ may be found numerically as that value that, for any given $q$, produces the required "equality" between $\pi^{(q)}_{po}$ and $\pi^{(q)}_{po[G]}$. This course has been followed here.

Section 5.1.2 identified the minCV and 1se models as significant special cases. The associated lambda values are denoted $\lambda'^{(minCV)}_{[G]}$ and $\lambda'^{(1se)}_{[G]}$ respectively. Subsequent sections will rely mainly on these as relating to reasonable priors (equivalently, selections of $\lambda'_{[G]}$, that may not be specific to any LASSO model $q$). Moreover, as each of these models will usually exhibit a high degree of compatibility with the data, the posterior distribution that follows from its prior will usually be centred somewhere in the vicinity of the posterior mode. In fact, the 1se model will be defined as the **primary model**, meaning that the actuary is assumed to use this as the source of an adopted loss reserve.

However, it will also be useful to investigate the extent to which $\lambda'_{[G]}$ may credibly deviate from these selections. Therefore, two values of this dispersion parameter are identified as of special interest. These may be labelled "simple-model" and "complex-model" dispersion parameters, and are defined as follows:

- **Simple:** the parameter is $\lambda'_{[G]} = \lambda'^{(simp)}_{[G]}$ such that;

$$\sum_{q=q_{1se}}^{Q} d\pi_{po[G]}\left(M_q|y, \lambda'^{(simp)}_{[G]}\right) = \varepsilon; \qquad (5.11)$$

and

- **Complex:** the parameter is $\lambda'_{[G]} = \lambda'^{(comp)}_{[G]}$ such that;

$$\sum_{q=1}^{q_{min}} d\pi_{po[G]}\left(M_q|y, \lambda'^{(comp)}_{[G]}\right) = \varepsilon; \qquad (5.12)$$

for some suitably small value $\varepsilon > 0$.

Thus, $\lambda'^{(simp)}_{[G]}$ is the prior dispersion parameter which, when applied to all models $M_q$, is just high enough to force the aggregate prior probability of all models $M_{1se}$ and more complex below a threshold probability. This is the value of $\lambda'_{[G]}$ that, when applied across all $M_q$, generates models so simple that are even remotely consistent with the data and model $M_{1se}$. The parameter value $\lambda'_{[G]} = \lambda'^{(comp)}_{[G]}$ has a corresponding meaning; it selects out the most complex models acceptable. The threshold value used here is $\varepsilon = 0.0005$, so a good deal of liberty has been allowed in the range $\left(\lambda'^{(comp)}_{[G]}, \lambda'^{(simp)}_{[G]}\right)$.

Sometimes no value $\lambda'^{(simp)}_{[G]}$ or (particularly) $\lambda'^{(comp)}_{[G]}$ can be found satisfying the required condition (5.11) or (5.12). In this case, there is no member of the family of LASSO models so simple (or complex) that it is inconsistent with the data to the stipulated threshold.

Examples of these simple-model and complex-model posteriors for Data Set 4 are given in Figure 5-1, where values of $q$ are listed on the horizontal axis. The top half of the figure illustrates the simple model posterior probability, and the bottom half the complex. The minCV and 1se models are also marked on the horizontal axis.

**Figure 5-1 Examples of simple-model and complex-model posterior distributions**

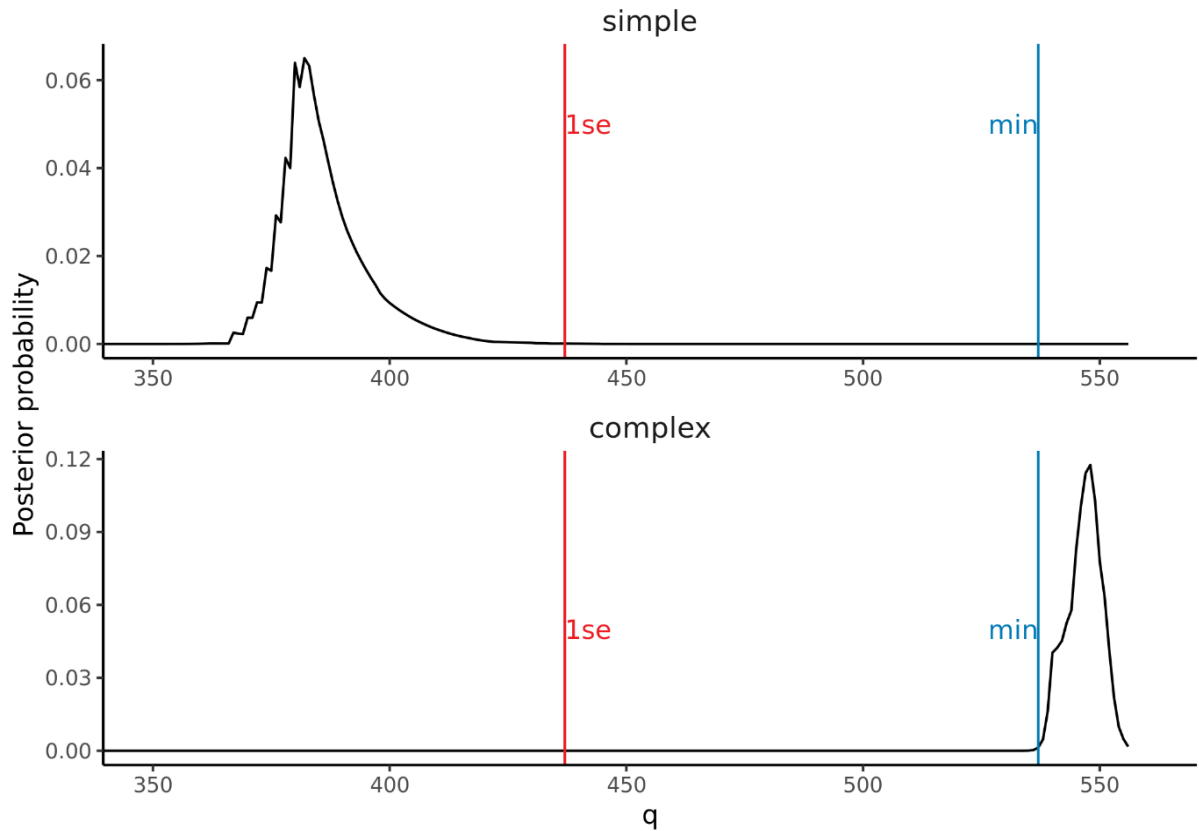

### 5.2.4. Prior distributions for Bayesian model averaging

The LASSO generates a family of models $\mathcal{F} = \{M_q, q = 1, \ldots, Q\}$, each tagged with a prior distribution characterized by dispersion parameter $\lambda'^{(q)}_{[G]}$. The prior relates to the set of possible parameters of the model $M_q$.

The aim is now to conduct BMA over $\mathcal{F}$. For this, a single prior is required across all models in $\mathcal{F}$. This might reasonably be taken as defined by a sensibly selected value $\lambda'_{[G]}$. For example, the selection $\lambda'_{[G]} = \lambda'^{(1se)}_{[G]}$ would take the prior distribution associated with LASSO model $M_{1se}$, and assume it to apply to all models in $\mathcal{F}$. When normalized across $\mathcal{F}$, this gives a prior distribution on $\mathcal{F}$, thus

$$Prob[M = M_q] = P_q = \frac{p_{1se}(\beta^{(q)})}{\sum_{q=1}^{Q} p_{1se}(\beta^{(q)})}, \qquad (5.13)$$

where $p_{1se}(\beta^{(q)})$ is obtained from (5.10) with the substitutions $\lambda = \lambda^{(1se)}_{[G]}, \beta = \beta^{(q)}$, this last term representing the parameter vector of model $M_q$.

BMA of any forecast of any random quantity $F$ then gives an average of

$$\bar{F}_{BMA} = \sum_{q=1}^{Q} P_q F_q, \qquad (5.14)$$

where $F_q$ is the forecast of $F$ given by model $M_q$, as parameterized by the LASSO.

Though it may be reasonable, as suggested above, to select the prior $p_{1se}(\beta_{1se})$ for use in (5.13), it is of interest to consider other possibilities. Therefore, re-write (5.13) in the more general form:

$$Prob[M = M_q] = P_q = \frac{p_0(\beta^{(q)})}{\sum_{q=1}^{Q} p_0(\beta^{(q)})}, \qquad (5.15)$$

where $p_0(\beta^{(q)})$ is constructed as for $p_{1se}(\beta^{(q)})$ except for the choice $\lambda = \lambda_{[G]}^{(0)}$, which is some preferred but unspecified prior.

BMA will be carried out below under various selections of prior, namely (in order of increasing model complexity):

- **Simple:** $\lambda_{[G]}^{(0)} = \lambda_{[G]}^{(simp)}$;
- **1se:** $\lambda_{[G]}^{(0)} = \lambda_{[G]}^{(1se)}$;
- **minCV:** $\lambda_{[G]}^{(0)} = \lambda_{[G]}^{(minCV)}$;
- **Complex:** $\lambda_{[G]}^{(0)} = \lambda_{[G]}^{(comp)}$.

The middle two of these would be regarded as yielding realistic results. The other two indicate the extremes to which results may be taken before consistency with data is totally lost.

### 5.2.5. Pruning of model set

Loss reserve forecasting is essentially an exercise in extrapolation. Certain functions are fitted to past data, and then extrapolated to forecast future data.

In the highly supervised environment of conventional loss reserving, this rarely creates difficulty. The actuary maintains strict control over the explanatory functions, often assigning physical explanations to them. Supervision of extrapolation behaviour is part of this control; if it appears unreasonable, the actuary will modify the model.

In the case of the LASSO, and other machine learning methods, on the other hand, the volume of models under test precludes this level of supervision. Moreover, the functions used to fit past data are assembled abstractly from a set of basis functions. Interpretation of the results may or may not be elementary.

It is in the nature of such exercises that strict control may be maintained over the fit of models to the data (e.g. by cross-validation error, as in Section 5.1.2), yet there be no control over the behaviour of these models in future. Hence, good quality fit of a model to data provides no assurance of reasonable forecast behaviour.

Figure 5-2 provides an example of this in relation to Data Set 2. Claim payments, aggregated over the last 10 accident quarters, have been plotted in $B against payment quarter for both

past and future quarters, in respect of both the primary model and an alternative candidate. The past includes payment quarters up and including $t = 40$, as indicated in the figure. The future is represented by $t > 40$.

**Figure 5-2 Example of high-quality fit to data but poor extrapolation**

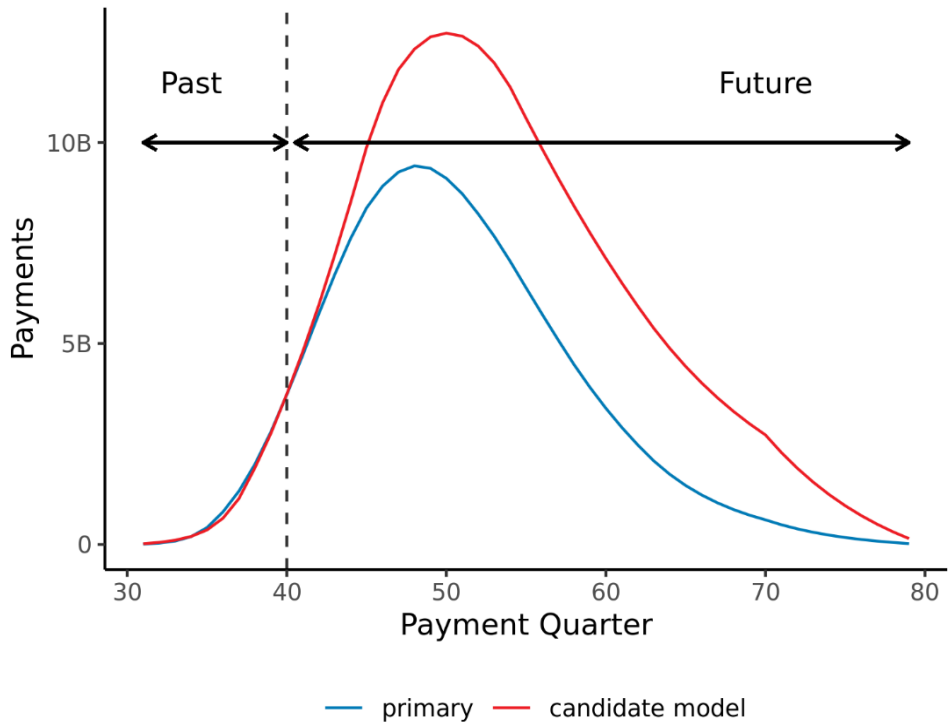

It can be seen that the candidate model coincides with the primary almost perfectly over the past payment quarters, and its posterior probability is in fact 0.67. However, its forecast quickly diverges from the primary's over future payment quarters. Ultimately, it forecasts future liability (for the last 10 accident quarters) 56% higher than the corresponding primary forecast.

It is highly unlikely that an actuary, having invested in a primary model, would regard a model that diverges to this degree as suitable for inclusion in the model set. If surveillance over inclusions were maintained, this particular case would be censored. For practical purposes, this censorship must be automated.

Table 5-1 parameterizes the pruning used here. The forecasts of each model are confronted by a number of **inclusion gates**. If they fail to pass through all of these gates, the model is censored.

**Table 5-1  Pruning of LASSO models**

| Component of forecast | Inclusion gates (as percentage of primary forecast) | |
|---|---|---|
| | **Lower limit** | **Upper limit** |
| **Accident quarters** | | |
| Last 2 | 75 | 133 |
| Last 5 | 80 | 125 |
| Last 10 | 83 | 120 |
| **Payment quarters** | | |
| Last 2 | 91 | 110 |
| Last 5 | 87 | 115 |
| Last 10 | 83 | 120 |

It is found that the dominant mode of censorship relates to breaches of the accident, rather than payment, quarter limits. This seems unsurprising as the limited experience of recent accident years provides little guide as to future development. It is easy for extrapolation functions to go awry.

The inclusion gates must be subjectively chosen, since they represent the actuary's judgement on the maximum acceptable deviation from the primary model. The gates are likely to vary from one reserving exercise to another, as the line of business, length of development tail, and other particulars change.

It is noteworthy that this introduces an unwelcome element of subjectivity into an otherwise data-driven estimation process. Unfortunately, we do not see an alternative. Failure to address the issue causes delinquent models to lead to unduly large estimates of internal model error that lack credibility.

### 5.2.6. Internal model structure error from the primary model

Equations (4.1) and (4.2) give expressions for the mean and variance of a posterior distribution of loss reserve. Equation (4.4) is the counterpart of (4.1) when the posterior derives from BMA across LASSO models. It will often be convenient to convert the posterior variance to a coefficient of variation **("CoV")**.

In this context, the posterior mean may be interpreted as an estimate of the loss reserve, and the CoV as a measure of IMSE. Numerical examples of this are presented in Section 7.2.

## 6. Bootstrapping the LASSO

### 6.1. Motivation

Section 5.2.6 described the estimation of IMSE for the primary model. In this sense, the problem is solved at that point. However, the estimate may be regarded as a little "thin" since the number of models with posterior probability greater than vanishingly small may be limited. It would not be unusual for this number to be in the range 10-30. This is likely to lead to under-estimation of the IMSE.

Bootstrapping of the LASSO generates multiple pseudo-data sets, leading to multiple posterior distributions, and adding to the reliability of the IMSE estimate.

### 6.2. Form of bootstrap

A semi-parametric form of bootstrap is used. This terminology is taken from Shibata (1997), and its implementation in relation to loss reserving (in relation to a GLM in that case) is explained in Taylor and McGuire (2016).

Pseudo-data sets are formed on the basis of the LASSO standardized residuals. Standardization requires an estimate of variance for each cell. In accordance with the assumption of a Gamma distribution in Section 5.1.2, the variance of the $(i,j)$ cell is estimated as $\hat{\phi}\hat{\mu}_{ij}^2$, where $\hat{\mu}_{ij}$ is the value fitted to the cell by the LASSO, and $\hat{\phi}$ is the estimate of $\phi$ obtained as described at the end of Section 5.1.2.

The LASSO is known to be biased, and so the set of standardized residuals must be centred before use in the bootstrap. They are all shifted by a constant quantity such that their arithmetic average is zero. One needs to check that the resulting centred standardized residuals appear iid, and they are then re-sampled to generate multiple pseudo-data sets $Y^{[b]}, b = 1, \ldots, B$ in the usual way, i.e. $Y_{ij}^{[b]} = \rho_{ij}^{[b]}\hat{\mu}_{ij} + \hat{\phi}\hat{\mu}_{ij}^2$, where the $\rho_{ij}^{[b]}$ are the re-sampled residuals. Hereafter, all quantities superscripted by $[b]$ will relate to this $b$-th bootstrap replication.

Each pseudo-data set is then submitted to the entire process described in Section 5, i.e. LASSO estimation, followed by BMA, with some modifications to censor unreasonable models and to centre the bootstrapped results to the original forecasts for each prior. Centring the bootstraps leads to a change in posterior probabilities so the process followed is:

- Carry out LASSO estimation to produce models $M_q^{[b]}, q = 1, \ldots Q^{[b]}$, their liability estimates $\hat{R}\left(M_q^{[b]}|Y^{[b]}\right)$, and associated posterior probabilities $d\pi_{po}^{[b]}\left(M_q^{[b]}\right)$. Note that it may be advisable to use a less onerous level of censoring at this level than set out in Table 5-1 to avoid eliminating models that may be acceptable after centring (next point). Hence temporary inclusion gates are obtained by scaling those in the table by a factor of 1.4.
- Calculate the scaling factor required to scale the sample mean of the posterior means from each pseudo data set to the original posterior means.
- Apply this scaling factor to the estimates of each model and recalculate the liability estimates, $\hat{R}\left(M_q^{[b]}|Y^{[b]}\right)$.
- Apply censoring at the desired level as discussed in Section 5.2.5.
- Calculate the associated posterior probabilities $d\pi_{po}^{[b]}\left(M_q^{[b]}\right)$.

- Drop any bootstrap replication $b$ for which posterior distribution over models $M_q^{[b]}$ is concentrated on less than 5 models in the sense of having fewer than 5 models with posterior probability masses each exceeding 0.0001.

The subsequent BMA is dealt with in Section 6.3.

6.3. The bootstrap matrix

Now form a $B \times Q_{max}$ matrix (**the bootstrap matrix**), where $Q_{max} = max\{Q^{[b]}, b = 1, \ldots, B\}$, and the $b$-th row contains the liability estimates $\hat{R}\left(M_q^{[b]}\right)|Y^{[b]}$, and contains missing values for columns $q = Q^{[b]} + 1, \ldots, Q_{max}$. Some of the rows of the matrix will be empty when all models of those rows have been censored. In machine learning parlance, the bootstrap matrix might be considered an ensemble, or a **ragged array**.

Note that, for fixed $q$, the models $M_q^{[b]}$ vary over $b$, i.e. a given column of the matrix might not contain the same model in all rows. Indeed, the models might be different in all rows of that column.

BMA produces an estimated liability $E_{po}\left[\hat{R}^{[b]}(M)|Y^{[b]}\right]$ and estimated IMSE equal to $CoV_{po}\left[\hat{R}^{[b]}(M)|Y^{[b]}\right]$ for data $Y^{[b]}$, where the subscripts $po$ indicate that the moments are taken with respect to the posterior distribution (compare with Section 4.3). Denote these quantities by $m^{[b]}$ and $w_{ISME}^{[b]}$ respectively. In addition, let $s_{IMSE}^{2[b]} = \left(m^{[b]} w_{ISME}^{[b]}\right)^2$.

Impute a uniform distribution $U$ across the rows of the matrix, i.e. all bootstrap replications are considered equally likely. Then more reliable estimators of liability and IMSE can be obtained by averaging over rows of the bootstrap matrix. These are respectively

$$m = E_U\left[m^{[b]}\right] = E_U E_{po}\left[\hat{R}^{[b]}(M)|Y^{[b]}\right], \tag{6.1}$$

$$s_{IMSE}^2 = E_U\left[s_{IMSE}^{2[b]}\right] = E_U Var_{po}\left[\hat{R}^{[b]}(M)|Y^{[b]}\right], \tag{6.2}$$

$$w_{IMSE}^2 = CoV_{IMSE}^2 = \frac{s_{IMSE}^2}{m^2}. \tag{6.3}$$

Conventional use of the bootstrap in loss reserving enables estimation of PaE. This is achieved by means of a well-established procedure (England and Verrall, 1999; Taylor and McGuire, 2016). PaE is estimated from the variation in loss reserve estimates over bootstrap replications. Process error can be added quite separately, and does not require the bootstrap.

A parallel procedure can be followed in relation to the bootstrap matrix here. One can calculate

$$s_{Pa}^2 = Var_U\left[m^{[b]}\right] = Var_U E_{po}\left[\hat{R}^{[b]}(M)|Y^{[b]}\right], \tag{6.4}$$
$$w_{Pa}^2 = \frac{s_{Pa}^2}{m^2}$$

and this is an estimate of parameter error. Once again, the estimation of process error CoV, denoted $w_{Pr}$, is separate and straightforward, using sampling from its known distribution (Section 6.2), as described by England and Verrall (1999, 2001, 2002), Taylor (2000) and Taylor and McGuire (2016) .

Note that, by the law of total variance,

$$Var[\hat{R}^{[b]}(M)] = Var_U E_{po}[\hat{R}^{[b]}(M)|Y^{[b]}] + E_U Var_{po}[\hat{R}^{[b]}(M)|Y^{[b]}] = s_{Pa}^2 + s_{IMSE}^2, \quad (6.5)$$

by (6.2) and (6.4), where the Var and E operators are understood as forming summary statistics from the sample values $\hat{R}^{[b]}(M)$, and where the variance on the left has been taken over the entire bootstrap matrix, i.e. over rows and columns.

The combination of PaE and IMSE has squared CoV

$$w_{Pa+IMSE}^2 = \frac{s_{Pa}^2 + s_{IMSE}^2}{m^2} = w_{Pa}^2 + w_{IMSE}^2. \quad (6.6)$$

**Remark 6.1.** The thinness of the primary estimate of IMSE, and the possibility of under-estimation, was noted in Section 6.1. Each bootstrap replication has the same stochastic properties as the primary, and so each row of the bootstrap matrix, and also the average of them, may lead to under-estimation. ∎

**Remark 6.2.** In view of Remark 6.1, the meanings of IMSE and PaE, as estimated by (6.2) and (6.4), merit some further consideration. The meaning of the latter is quite clear in the case of a GLM. Here there is a fixed model structure across all bootstrap replications, and it is applied to the various data sets generated by those replications. Only the parameterization of that structure varies across replications. Quite clearly, this leads to genuine estimation of PaE.

When the LASSO is bootstrapped, the situation is quite different. Variation of the data set from one replication to another, induces different sets of models. Hence, not only parameterization of models, but the models themselves, vary across replications. In this way, parameter and model errors become entangled, and the distinction between them becomes unclear. Indeed, there may even be a question as to whether the distinction is even meaningful.

From a practical viewpoint, it may be preferable to think in terms of just (6.6) as unambiguously the combination of parameter and IMSE errors without concern for the decomposition into the two components. For most practical purposes (e.g. risk margins), only the combination will be required.

The estimates $w_{ISME}$ and $w_{Pa}$, as defined by (6.2) and (6.4), are useful for the construction of (6.6), but no particular meaning may be assigned to the components themselves. ∎

# 7. Numerical results
## 7.1. Data sets
All data submitted to analysis in this paper are synthetic. Thus, all data sets contain known features. Four synthetic data sets were constructed and analyzed. Each consisted of a $40 \times 40$ quarterly triangle of incremental paid claims $Y_{ij}$; thus 810 cells. The data were simulated according to exactly the same specifications as set out in McGuire, Taylor and Miller (2021).

Full details were given in that earlier paper. It suffices here to give just the briefest descriptions of the data sets, as below.

**Data set 1:** Cell means consist of multiplicative row and column effects, and so this data set is a simple one, consistent with a chain ladder model.

**Data set 2:** Cell means are as for Data set 1, except that a payment quarter effect has been added. This effect represents SI at a rate that is constant over the first 12 quarters, increases over the next 12, returns to constant over the next 8, then increases once more over the final 8. For any one payment quarter, the rate of SI is constant across development quarters.

**Data set 3:** This is as for Data set 2, except that one further complexity has been added. This consists of a sharp increase in claim costs in development quarters 21 and later within accident quarters 17 and later. This last feature is of particularly difficulty for detection by a model, as it affects only 10 of the 810 cells of which the data set is composed.

**Data set 4:** This is as for Data set 2 except that greater complexity has been added to the rates of SI. The rates from Data set 2 now apply to development quarter 1 and, for any payment quarter, they decline linearly to zero at development quarter 40.

These data sets might reasonably be listed 1, 2, 3, 4 in increasing order of complexity, though estimation might be more difficult in case 3 than case 4.

## 7.2. Primary model

Each of the four data sets of Section 7.1 are modelled and extrapolated to produce an estimated loss reserve and IMSE on the basis of the primary model, as described in Section 5. The results are set out in Table 7-1. Since the data are synthetic, the true reserves are known, and these have also been included in the table.

The table includes two sets of forecasts. The "raw 1se" is the 1se estimate from the primary model, obtained by a single application of the LASSO to the data set. The "posterior" is the mean of the posterior distribution obtained in the BMA described in Section 5.1.2.

The "1se" and "minCV" rows are bolded as the most relevant. The "simple and "complex" rows are also included to indicate the extent to which IMSE may vary before the models taken into account by it exhibit lack of fidelity to the data.

It may be noted that the posterior means usually exceed the true liabilities. This is not a product of the BMA, as the difference between the raw 1se estimate and its Bayesian average is small in all cases. The 1se models simply over-estimate liability in these cases.

**Table 7-1  Forecast reserve and estimated IMSE for primary models**

| Data set | LASSO model | Loss reserve | | | Estimated IMSE (CoV) |
|---|---|---|---|---|---|
| | | True | Forecast | | |
| | | | Raw 1se | Posterior | |
| | | $B | $B | $B | % |
| 1 | Simple | 190 | | 198 | 0.7 |
| | **1se** | **190** | **194** | **194** | **0.4** |
| | **minCV** | **190** | | **194** | **0.5** |
| | Complex | 190 | | 203 | 0.8 |
| 2 | Simple | 238 | | 260 | 0.1 |
| | **1se** | **238** | **261** | **260** | **0.1** |
| | **minCV** | **238** | | **244** | **3.4** |
| | Complex | 238 | | 272 | 2.9 |
| 3 | Simple | 608 | | 877 | 1.7 |
| | **1se** | **608** | **778** | **777** | **6.8** |
| | **minCV** | **608** | | **687** | **2.0** |
| | Complex | 608 | | 874 | 5.5 |
| 4 | Simple | 216 | | 244 | 0.2 |
| | **1se** | **216** | **247** | **247** | **0.3** |
| | **minCV** | **216** | | **268** | **0.7** |
| | Complex | 216 | | 276 | 1.2 |

The modelling of Data set 1 behaves well, with small estimates of model error, as is to be expected since the data is compatible with a simple chain ladder model, reflecting just row and column effects.

It is to be emphasized that, although Table 7-1 focuses attention on the CoV of IMSE, the full posterior distribution of loss reserves is available for each row of the table. As an example, Figure 7-1 displays the posterior corresponding to the 1se prior in the case of Data set 2. The convolution of this with the distributions of other components can be taken to obtain the distribution of full forecast error.

**Figure 7-1 Posterior distribution of loss reserve for 1se prior in Data set 2**

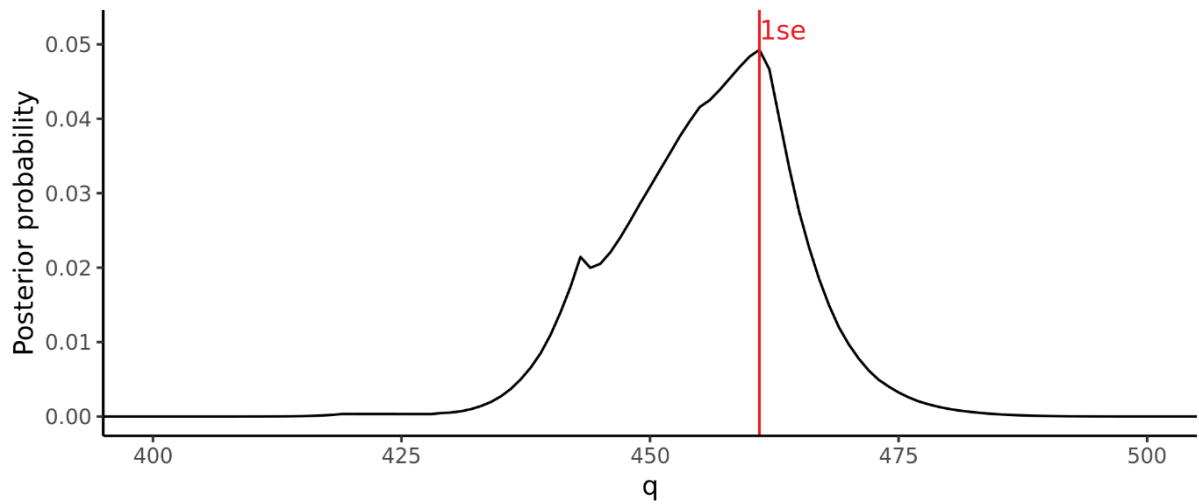

## 7.3. Bootstrap

The bootstrap procedure described in Section 6 was applied to all four data sets. In each case, the number of bootstrap replications was set to $B = 200$, although, as will be seen in the results below, some of these are discarded by the pruning regime of Section 5.2.5. Table 7-2 displays the results corresponding to the primary model results in Table 7-1.

**Table 7-2  Forecast reserve and estimated IMSE for bootstrapped models**

| Data set | LASSO prior | Loss reserve | | Estimated IMSE (CoV) | Number of surviving bootstraps |
|---|---|---|---|---|---|
| | | True | Posterior forecast | | |
| | | $B | $B | % | |
| 1 | Simple | 191 | 200 | 2.0 | 171 |
| | **1se** | **186** | **197** | **1.9** | **169** |
| | **minCV** | **185** | **196** | **2.4** | **163** |
| | Complex | 197 | 206 | 2.2 | 162 |
| 2 | Simple | 238 | 253 | 2.0 | 180 |
| | **1se** | **238** | **252** | **2.0** | **179** |
| | **minCV** | **238** | **240** | **2.5** | **147** |
| | Complex | 238 | 251 | 2.4 | 148 |
| 3 | Simple | 608 | 787 | 2.4 | 78 |
| | **1se** | **608** | **703** | **2.8** | **75** |
| | **minCV** | **608** | **589** | **2.7** | **66** |
| | Complex | 608 | 720 | 2.3 | 77 |
| 4 | Simple | 216 | 241 | 1.8 | 168 |
| | **1se** | **216** | **243** | **1.9** | **165** |
| | **minCV** | **216** | **252** | **2.6** | **174** |
| | Complex | 216 | 258 | 2.2 | 170 |

A comparison of Table 7-2 with Table 7-1 reveals the effect of bootstrapping in supplementing the primary model. The magnitudes of these estimates of IMSE vary over the four data sets somewhat as expected; at least, the largest estimate occurs for the exotic Data set 3. It is interesting that there is little difference between the estimates for Data sets 1, 2 and 4, even though the complexity varies between them. This suggests that the LASSO models perform about equally well in these three cases.

The final column illustrates how the censorship rates increase with increasing complexity of the data set. In other words, as models are confronted with increasing data complexity, those models include additional feature that render their extrapolation more prone to delinquency. This is particularly the case for Data set 3, for which the forecasting difficulties have been discussed in Section 7.1.

As has been discussed in Section 6.3, the LASSO bootstrap provides an estimate of PaE in addition to ISME, and process error may be estimated routinely. These are displayed in Table 7-3. Only the cases of 1se and minCV priors are included as those of greatest interest.

As noted in Remark 6.2, there is some ambiguity in the separate meanings of PaE and IMSE, but not in the meaning of their combined effects. Table 7-3 therefore also contains a column displaying the CoV associated with the sum of these two components of forecast error.

The "Sub-total" error is the sum of parameter, process and IMSE. It is designated in this way to emphasize that it is not, in fact, the total forecast error because it omits the two components not discussed in this paper, namely

- External model structure error; and
- Model distribution error.

**Table 7-3 Forecast reserve and estimated forecast error components for bootstrapped models**

| Data set | LASSO prior | Liability | | Forecast Error (CoV) | | | | |
|---|---|---|---|---|---|---|---|---|
| | | True | Posterior | IMSE | Parameter | IMSE + parameter | Process | "Sub-total" |
| | | $B | $B | % | % | % | % | % |
| 1 | 1se | 190 | 186 | 1.9 | 9.2 | 9.4 | 3.9 | 10.2 |
| | minCV | 190 | 185 | 2.4 | 10.5 | 10.8 | 4.4 | 11.6 |
| 2 | 1se | 238 | 252 | 2.0 | 10.0 | 10.2 | 3.9 | 10.9 |
| | minCV | 238 | 240 | 2.5 | 8.8 | 9.2 | 4.7 | 10.3 |
| 3 | 1se | 608 | 703 | 2.8 | 11.2 | 11.6 | 5.7 | 12.9 |
| | minCV | 608 | 589 | 2.7 | 11.2 | 11.5 | 5.3 | 12.7 |
| 4 | 1se | 216 | 243 | 1.9 | 8.6 | 8.8 | 4.0 | 9.7 |
| | minCV | 216 | 252 | 2.6 | 12.5 | 12.8 | 5.1 | 13.8 |

It may be noted that the magnitude of forecast "sub-total" error usually depends little on whether it is derived from a 1se or a minCV prior distribution. There is one exception to this, Data set 4, where the difference relates mainly to estimated "parameter error".

Investigation reveals that the minCV distribution of posterior means across bootstrap replications exhibits a markedly longer right tail than its 1se counterpart. The reasons for this have not been investigated further, but note the complex structure of SI for this data set. Mis-estimation of SI as a result of undue model complexity could lead to volatile forecast along future diagonals.

There is some support for this explanation in Table 7-4, which repeats the parameter error CoVs from Table 7-3, but now flanked by those resulting from the simple and complex priors. The table displays a clear divide between the two simpler and the two more complex cases.

**Table 7-4  Bootstrap estimates of parameter error for Data set 4**

| Prior | Parameter error (CoV) |
|---|---|
| | % |
| simple | 8.1 |
| 1se | 8.6 |
| minCV | 12.5 |
| complex | 13.4 |

### 7.4. Sensitivity to inclusion gate width

All numerical results reported above are governed by the inclusion gates set out in Table 5-1. Since these gates were subjectively determined, it is natural to enquire into the sensitivity of the results to the gate-widths.

Table 7-5 compares the "Sub-total" forecast errors of Table 7-3 with the errors that are obtained when all gates are widened by increasing the upper limit by a factor of 1.1, and decreasing the lower limit by the same factor. Thus the first gate in Table 5-1 is widened from [0.75,1.33] to [0.68,1.46], and the other gates in that table widened similarly.

**Table 7-5  Sensitivity of forecast error to width of inclusion gates**

| Data set | Prior | Sub-total error (CoV) | |
|---|---|---|---|
| | | Original gates | Widened gates |
| | | % | % |
| 1 | 1se | 10.2 | 14.9 |
| | minCV | 11.6 | 18.5 |
| 2 | 1se | 10.9 | 15.3 |
| | minCV | 10.3 | 17.7 |
| 3 | 1se | 12.9 | 18.0 |
| | minCV | 12.7 | 19.4 |
| 4 | 1se | 9.7 | 15.1 |
| | minCV | 13.8 | 19.6 |

It is seen that, in very broad terms, widening the inclusion gates by a factor of 1.1 increases forecast error by a factor of about 1.5. Thus, estimated forecast error can be sensitive to the selection of these gates, and they must be selected carefully.

## 7.5. Reasonableness check of results

It is interesting to use a different methodology in an attempt to provide an independent check on the reasonableness of the above estimates of PaE plus IMSE. For this purpose, a GLM has been fitted to each data set, treating the model structure as known, i.e. containing the same explanatory variables as in the simulation specifications. Thus, the GLMs should have low levels of IMSE. The GLMs have then been bootstrapped to yield estimates of PaE. The GLM bootstrap forecasts were subject to the same set of inclusion gates as applied to the LASSO (see Table 5-1).

Before examining the results of these models, it is useful to consider reasonable expectations of them. As noted in Remark 6.2, some IMSE may leak into recorded PaE in the case of the LASSO. Accordingly, it would not be surprising if the PaE, as estimated by the LASSO, exceeded that estimated by the GLM.

**Table 7-6  Comparison of LASSO and GLM estimates of parameter error**

| Data set | Prior | Forecast True | Forecast Mean | Parameter error (CoV) |
|---|---|---|---|---|
|  |  | $B | $B | % |
| 1 | LASSO 1se | 190 | 186 | 9.2 |
|  | GLM | 190 | 214 | 4.8 |
| 2 | LASSO 1se | 238 | 252 | 10.0 |
|  | GLM | 238 | 211 | 5.7 |
| 3 | LASSO 1se | 608 | 703 | 11.2 |
|  | GLM | 608 | 631 | 8.3 |
| 4 | LASSO 1se | 216 | 243 | 8.6 |
|  | GLM | 216 | 232 | 6.7 |

Table 7-6 provides the comparison. It is certainly the case that the LASSO 1se estimates of PaE exceed the GLM estimates. Specific observations are that:

- the very simple Data set 1 and the more complex Data set 2 both produce substantial differences;
- the discrepancies for Data sets 3 and 4 are also significant, but of lower magnitude.

These results appear consistent with the expectations set earlier in this sub-section.

# 8. Conclusions

Section 1 noted that this paper set out to study model error. This is all about forecasting under model uncertainty. Sections 5 and 6 have indeed set out a LASSO-based methodology for the estimation of a specific component of model error, namely internal model structure error, and the methodology has been illustrated numerically in Section 7.

With one exception, noted in Section 5.2.5, the estimation of IMSE outlined here is rigorous, based on Bayesian model averaging over a space of admissible models. This provides an alternative to the subjective estimation found in prior literature.

The exception discussed in Section 5.2.5 is concerned with the fact that some models that fit a data set extremely well may yet forecast abominably. In such cases, there is no apparent data-driven means of excluding the delinquent models. This matter has been dealt with by the establishment of subjectively chosen inclusion gates. Models are rejected as unrealistic if they produce certain results that fail to pass through these gates.

The selection of these gates would be based on the reserving actuary's professional expertise. In one sense, the need to rely on this element of subjectivity is unfortunate. However, the issue cannot be disregarded, as delinquent models have the potential to corrupt the model error estimation completely, and produce wildly inflated estimates.

The authors see no alternative to some form of subjective decision on the admissibility of individual models. There are, however, obvious risks associated with it in that under- (over-) estimation of model error will result from unduly strict (lax) selection of the gates. Moreover, the selection requires considerable care as the estimate of model error is materially sensitive to it (Section 7.4).

The suggested estimation procedure is supplemented by bootstrapping in Section 6. This yields a bonus in that it produces an estimate of PaE as well as IMSE. However, as noted in Remark 6.2, the supposed estimates of PaE and IMSE are entangled, probably inextricably. Possibly, there is not even any clear meaning to be associated with these separate components.

But all is not lost here. Although there are not reliable estimates of PaE and IMSE (whatever they may be), Remark 6.2 points out that the estimate of their combined contribution to forecast error is indeed reliable, and this is all that is required for most practical purposes, such as loss reserve risk margins.

The estimates of PaE and IMSE obtained from the LASSO have been subjected to a sanity check in Section 7.5 by comparison with the PaE estimate obtained from a GLM, and appear generally reasonable.

It is important to note that, as pointed out in Section 3, IMSE is but one component of model error. The estimation of IMSE, PaE and process error are discussed in this paper, and illustrated numerically in Section 7, but this leaves model distribution error and external model structure error still to be brought to account.

The investigations of Taylor (2021) hint that the first of these would usually be small, but the second often would not. The only discussion of this component of forecast error in the actuarial literature occurs in O'Dowd, Smith and Hardy (2005) and Risk Margins Task Force (2008), where a subjective approach is taken.

This may well be the only option in making allowance for influences that are totally divorced from past experience and data. Be that as it may, the fact remains that any estimate of external model structure error must be derived by means quite different from those used for IMSE. If a subjective estimate is required, then, once obtained, it may be combined with the other components dealt with in this paper.

# Acknowledgment

The research reported here received financial supported under Australian Research Council's Linkage Projects funding scheme (project number LP130100723). The authors also acknowledge valuable discussion with Gael Martin of Monash University, Benjamin Avanzi and David Yu, both of Melbourne University, and Bernard Wong of University of New South Wales.